# Underlying Structure-Activity Correlations of 2D Layered Transition Metal Dichalcogenides-Based Electrocatalysts for Boosted Hydrogen Generation

Zhexu Xi

*Bristol Centre for Functional Nanomaterials, University of Bristol, Bristol, UK*

**Abstract:** Hydrogen fuel is an ideal energy source to replace the traditional fossil fuels because of its high energy density and renewability. Electrochemical water splitting is also regarded as a sustainable, cleaning and eco-friendly method for hydrogen evolution reaction (HER), but a cheaper, earth-abundant and similarly efficient alternative to Pt as an HER catalyst cannot still be discovered. Recently, 2D Transition Metal Dichalcogenides (TMDs) are demonstrated to greatly enhance the HER activity. Herein, our work provides an insight into the recent advances in 2D TMDs-based HER following the composition-characterisation-construction guideline. After the background introduction, several research outputs based on 2D TMDs as well as the comprehensive analysis on the modulation strategies of 2D TMDs, for the purposes of increasing the active sites, improving the intrinsic activity and altering the electronic states. Finally, the future opportunities and challenges of 2D TMDs electrocatalysts are briefly featured.
**Key words:** electrocatalysts; transition metal dichalcogenides; hydrogen evolution; modification; basel plane; active sites

## 1. Introduction

Nowadays, demand for usable energy worldwide has dramatically risen due to rapid growth in population, which inevitably triggers the overuse of traditional fossil fuels as well as a series of environmental issues[1, 2]. Accordingly, it is of great importance to find another, less polluting energy source to tackle the current problems. Hydrogen ($H_2$), owing to its zero-polluting combustion byproduct (water) and high energy density, holds high potential as an alternative to fossil energy[3]. For $H_2$ production pathways, water electrolysis (electrocatalytic water splitting) is also known as a renewable and clean industrial approach[4]. Currently, the best electrocatalyst for the Hydrogen Evolution Reaction (HER) is Pt, which markedly minimizes the overpotential and exhibits optimal catalytic activity. However, the high cost and limited reserves of Pt seriously restrict the further development of Pt-based catalysts[3, 5]. Thus, a novel HER electrocatalyst with rich abundance and similar reactivity to Pt has captured wide attention.

Two-dimensional transition metal dichalcogenides (2D TMDs), also generally expressed in the form of $MX_n$ (M = Mo, W, Ti, V, and Zr; X= S, Se, and Te), have recently been verified to be the most prospective promising alternatives to Pt due to extraordinary catalytic performance[4-6]. First, the atomically thin 2D layered structure offers plentiful exposed active sites and a high specific area for HER[3][7]. Second, the unique characteristics of TMDs are primarily related to the tailor-made electronic structures, which can provide a more accurate and comprehensive understanding in terms of their HER catalytic mechanisms[7, 8]. Third, although the unsatisfactory in-plane activity of TMDs has been reported to restrict the applications in HER electrocatalysts, more strategies based on the structural modification of TMDs have been implemented to improve the catalytic performance, including phase transition and defect engineering[8, 9].

Herein, we focus on the role of 2D TMDs as ideal replacements for Pt in HER enhancement. First, we summarise the theoretical understanding of the overall electrocatalytic HER system. Second, based on the recent discoveries in this rapidly advancing research area, we make a comprehensive analysis regarding the nanoscale modulation strategies of 2D-TMDs-based electrocatalysts in three aspects: 1) composition (different kinds of 2D TMDs that have superior HER catalytic performance, as well as how structural modulation strategies are implemented), 2) characterisation (various instrumental techniques used for measurement, analysis and quantification of 2D TMDs), and 3) construction (novel nanoengineered 2D TMDs based on versatile modulation strategies that boost the HER activity)[3-5, 7-9, 10]. Finally, we propose the perspectives and challenges of TMDs-based electrochemical water splitting technologies, which can provide more insights into the rational design and fabrication of HER-related catalysts.

## 2. Fundamentals of HER

For HER electrocatalysis, three elementary reactions are involved, including one discharge step and two different hydrogen desorption steps (chemical and electrochemical pathways). As Fig. 1 illustrates, a transferred electron initially forms an adsorbed hydrogen atom (H*) with a combination of a proton on the active site of the electrode surface (the Volmer or discharge reaction step), and then generates $H_2$ by combination with another adsorbed atom (the Tafel step) or a proton from the solution (the Heyrovsky step), which depends on the kind of the electrode material[11].

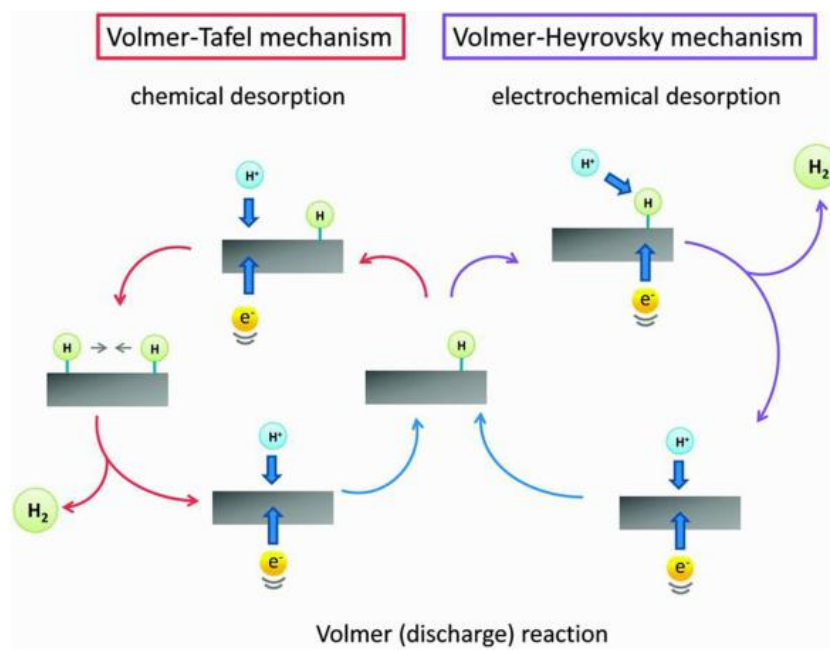

**Fig. 1** the mechanism of HER in acidic media[11].

Based on the HER process in acidic solutions, the critical parameter to thermodynamically control the entire catalytic reaction rate is the corresponding Gibbs Free Energy change for absorbed hydrogen atoms ($\Delta G_H$), which is connected with the balance between the H* adsorption and $H_2$ desorption step. From the physical chemistry perspective, if the hydrogen-catalyst bonding on the electrode surface is too strong, making the Volmer step more straightforward, the desorption reaction will be the rate-determining step; otherwise, the initial Volmer step will be started with a tremendous driving force and accordingly, the adsorption will be the rate-determining step[8]. Hence, a nearly zero value of $\Delta G_H$ can bring about the best HER performance, which has also become a constant focus of the optimal design of ideal TMD-based catalysts[8, 11].

## 3. General outline on the modification strategies of 2D TMDs for boosted HER performance

In the context of a comprehensive characterisation-composition-construction analysis, different categories of nanolayered TMDs have been firmly demonstrated to be ideal for HER enhancement due to their appealing electronic properties and similar stacking methods, so they are naturally the largest families of functionalised catalysts for electrochemical reactions[7, 12].

In the 1970s, layered TMDs have been identified as unsuitable electrocatalysts for hydrogen generation due to the low intrinsic activity in the bulk state[13]. Until 2005, Hinnemann et al.[14] found that the Mo-edge of $MoS_2$ (1010) is similar to the HER active sites of nitrogenase. Simultaneously, their calculation based on a Density Functional Theory (DFT) model shows a comparable $\Delta G_H$ value to Pt, which theoretically predicts the excellent HER performance. Then, they loaded $MoS_2$ nanoparticles on the graphite substrate to experimentally confirm their prediction. More recent research identifies that the edge of $MoS_2$ plays a more vital role than its inert basal plane in electrocatalytic HER, even revealing the linear relationship between the HER activity and the edge length[15]. Therefore, three main modification routes emerge: 1) structural engineering: for the limited edge sites of TMDs, to increase the concentration of edge sites by regulating the architecture on the nanoscale; 2) edge/in-plane intrinsic activity regulation: to significantly activate the inherently inert basal plane or edge sites; and 3) to optimise the electronic structure of TMDs.

The first route focuses substantially on increasing the number of exposed active sites; the second route is to lessen the charge transfer resistance; the third route is even more valuable for the overall HER catalytic activity enhancement.

## 4. Increasing the concentration of edge sites
### 4.1. Physical Exfoliation from bulk materials
Compared with bulk materials, the most appealing trait for 2D-layered TMDs is a relatively big specific surface area, which provides plenty of rooms for adequately exposed active sites on edge[10, 11].

To maximise the edge active sites in bulk state for enhanced HER activity, more well-tuned and various morphological features should be physically or chemically tuned. One direct and straight-forward physical method refers to thinning the thickness and numbers of layers. The typical approach is mechanical exfoliation from the bulk states to the monolayered TMD nanosheets. This easy way markedly contributes to the boosted HER activity, and more importantly, makes it easier for a better understanding of the entire HER experimental system. Accordingly, creating a TMD monolayer by exfoliation is widely acknowledged for the practical utility of the HER system. Using the imaging and analysis of exfoliated $MoS_2$ nanoplate as an example (Fig. 2), as illustrated in the scanning tunnelling microscopy (STM) result, the exfoliated $MoS_2$ nanoplate image provides a sufficiently clear image to graphically distinguish the morphological differences between the basal plane and the edge[16].

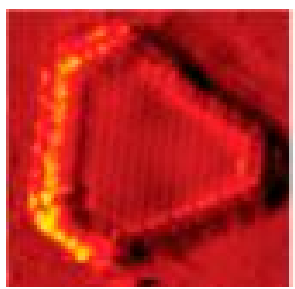

**Fig. 2** STM image of MoS2 nanoplate by exfoliation[16].

Moreover, the linear sweep voltammetry (LSV) measurement further indicates the significantly boosted HER performance after exfoliation with a milder Tafel slope of 94.31 mV/dec and lower onset potential, as shown in Fig. 3[17].

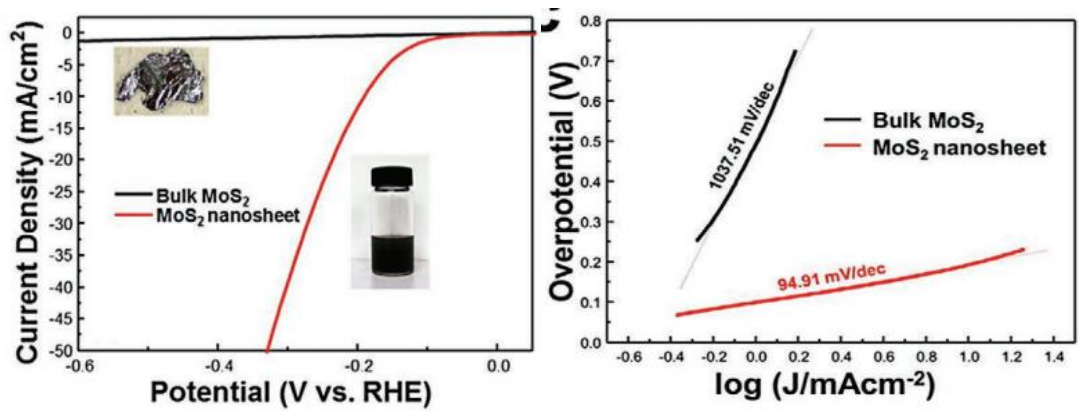

**Fig. 3** LSV curve (left) and the corresponding Tafel curve (right) between bulk MoS$_2$ and MoS$_2$ nanosheet to demonstrate the difference between both via chemical exfoliation methods[17].

## 4.2. Chemical formation of versatile nanostructured TMDs

Via chemical methodologies, various nanostructured morphologies can be well designed and constructed, like nanowires, nanosheets, nanoparticles, nanosized vertical alignments, mesoporous structures, and amorphous MoX$_2$[4, 10, 17, 18, 19]. These strategies strikingly reduce the overuse of raw materials under physical exfoliation. Zhang et al.[18] successfully generated high-density MoS$_2$ nanosheets vertically on polyaniline (PANI) nanowires, and synthesised 3D layered MoS$_2$/PANI composites. In this structure, PANI serves as architectural and electrical-conductive support for MoS$_2$, and the vertically aligned structure also exposes more edge active sites. MoS$_2$/PANI alignments deliver better HER catalytic performance than purely bulk states by maximising the most active edges[18].

Additionally, other useful morphological formations and characterisation techniques positively impact the overall analysis of versatile nanostructured TMDs. Binary TMDs, generated by incorporating another metallic atom into transition metal sulfides, suggest an extraordinarily enhanced activity compared with bulk materials. Li et al.[19] fabricated Mo$_{(1-x)}$-W$_x$-S$_2$ composite as an ideal HER catalyst by a mild hydrothermal synthetic method. Here, the scanning electron microscopy (SEM) images (Fig. 4) indicate a well-defined spherical architecture with tensely attached nanopetals. Morphologically, it coincides with the pristine structure. Also, the transmission electron microscopy (TEM) image gives a higher-resolved picture of 2D stacked nanopetals. For further analysis of the composition, the exact formula of the composite is Mo$_{0.85}$W$_{0.15}$S$_2$ by X-ray photoelectron spectroscopy (XPS).

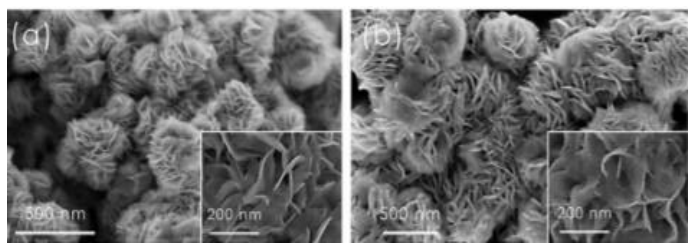

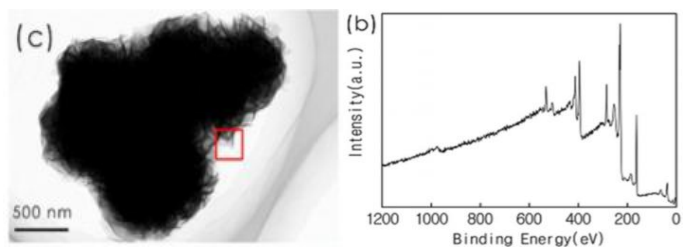

**Fig. 4 (a)** SEM image of the pristine MoS$_2$; **(b)** SEM image of the MoWS$_2$ composite; **(c)** TEM image of the MoWS$_2$ composite; **(d)** overall XPS spectra of the composite[19].

## 5. Regulating the intrinsic activity

### 5.1. Phase transition engineering

As one of the well-investigated TMD-related electrocatalytic materials, the activity of MoS$_2$ has been thoroughly understood. Due to its S-Mo-S layers in three different stacking ways, MoS$_2$ owns three possible prisms: 1T-, 2H-, and 3R-types[10]. 2H-MoS$_2$ is thermodynamically stable and exhibits the semiconducting characteristic, while the metastable 1T phase shows the metallic property. Consequently, the operable phase transition engineering from 2H to 1T makes a great difference to the electron conduction, driving the HER kinetics[8, 9, 12, 20]. Usually, 1T-types contribute to the boosted catalytic performance due to the higher density of active sites and metallic conductivity.

### 5.1.1. Ion intercalation

Lithium-ion intercalation is a common method for phase transformation. According to the test result of Lukowski et al.[20], the nanolayered 1T MoS$_2$ sheets reveal superior conductivity based on the AFM lithography in constant current mode. Similarly, from the electrochemical impendence spectroscopy (EIS) pattern, the charge transfer resistance of 2H sheets (232 Ω) sharply descends to 4 Ω with the transition, which also signifies a drastically rising electron conductivity (Fig. 5(b) and (c)). Likewise, Wang et al.[21] constructed a vertically arrayed electrode and precisely tuned the 2H-1T transition process by controlling the potentials, as Fig. 5(a) describes. More persuasively, their result explicitly reveals that the deeper Li$^+$ discharge process leads to more spacing expansion of MoS$_2$ nanosheets and then boosts the HER kinetics.

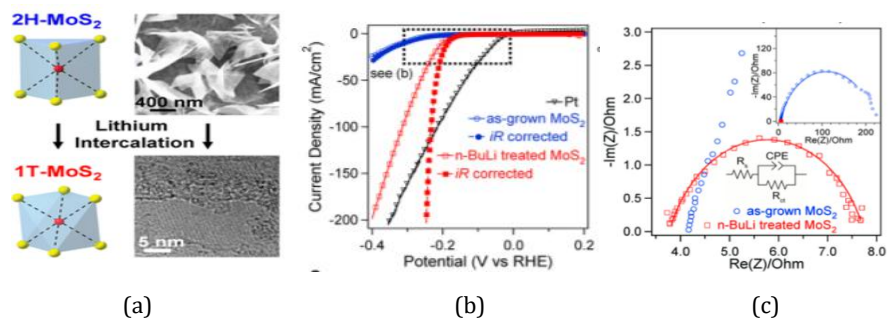

(a)          (b)          (c)

**Fig. 5 (a)** Schematic diagram of the phase transition process from 2H- to 1T-MoS$_2$ induced by Li$^+$ intercalation; **(b)** current-potential curve depicting the electrocatalytic performance of Pt, 1T- (n-BuLi treated, in red) and 2H-MoS$_2$ nanosheets (as-grown, in blue) fabricated under different pretreatment conditions; **(c)** comparison of as-grown and n-BuLi treated MoS$_2$ in EIS spectrum[21].

Intercalation of extra ions has two roles in the phase transition process: 1) introducing new metal cations changes $\Delta G_H$; 2) intercalation contributes to the electron injection into the nanosheets with the reduced charge and facilitates more catalytically active sites. With more intercalated ions, the intrinsic electronic structure can be efficiently and accurately modulated[22, 23].

### 5.1.2. Lattice strain

Besides ion intercalation, introducing strain also plays an indispensable role in the phase transition of TMDs. Tan et al.[24] synthesised a monolayered $MoS_2$ thin film on nanoporous Au substrate with a distorted surface. The surface distortion is demonstrated to form the lattice deformation and exert a strain-specific effect on the film: causing a measurable change in the S-Mo-S bond angle, thereby driving the localised 2H-1T phase transition. This approach can improve the state of density near the Fermi energy level by generating the corresponding stress, thereby leading to the localised 2H-1T phase transition and the loss of adsorption energy.

Recently, the combinations of atomic layer deposition (ALD) and electro-etching techniques were utilised to analyse the strain-affected HER performance. Specifically, Titanium Dioxide ($TiO_2$) was deposited on the plane of $MoS_2$ nanosheets by ALD, and the nanosheets were activated via an in-situ electrochemical method and then leached out to produce the strain. The testing result reveals a tremendous boost in the HER activity, as demonstrated in the LSV (Fig. 6(a)) and Tafel curves (Fig. 6(b)) respectively, and a positive correlation between the Tafel slope and the ALD cycle number (Fig. 6(c))[25].

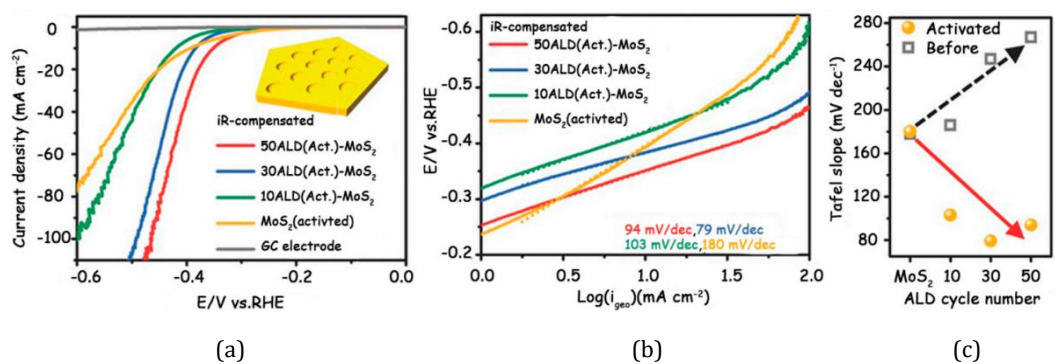

(a) (b) (c)

**Fig. 6 (a)** iR-compensated LSV curves of $MoS_2$ (activated) and ALD(act.) $MoS_2$ at different scan cycles; **(b)** Tafel curves made based on (a); **(c)** graph showing the dependence of the Tafel slope and the ALD cycle number[25]

### 5.2. Defect engineering

Apart from increasing the density of electrocatalytically active edge sites, the basel plane possesses a larger surface area and proportion of potential sites for HER. Intentionally creating a new structural defects on the nanostructures of TMDs can tune the density of states more precisely and efficiently activate the intrinsically inert plane, then regulating the value of $\Delta G_H$ in $H_2$ generation[4, 8, 11-13]. The common form of the introduced defects in nanolayered TMDs contains point defects (vacancy, and atom impurity) and line defects (grain boundaries, GBs).

### 5.2.1. Introducing vacancies

Introducing the moderate number of S-vacancies can significantly enhance the HER activity by creating additionally new in-plane active sites. Various techniques have been used to introduce more active vacancies, such as plasma treatment, chemical liquid exfoliation[26]. Compared with the LSV curves of pristine and plasma-treated $MoS_2$ with different time, with 15 minutes plasma treatment, $MoS_2$ has the optimal HER activity (Fig. 7), with the overpotential of 183 mV notably lower than that of pristine one (727 mV in Fig. 7(a))[26].

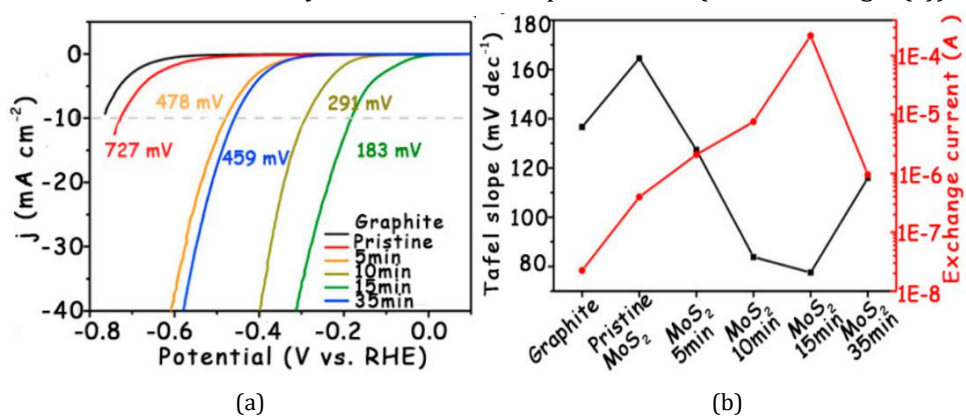

(a)  (b)

**Fig. 7 (a)** polarisation curves of plasma-treated $MoS_2$ with various times; **(b)** the corresponding Tafel curve and exchange current curve[26].

Several computational discoveries based on synergic tuning of S atom vacancies and strain demonstrates the HER activity decreases with the lower concentration of S-vacancies[27]. The corresponding simulation shows that the best HER activity and kinetics should be under the optimal percentage of vacancies and elastic strains[28]. Hence, these results provide an instructive suggestion for the well-tuned HER system by enriching S-vacancies and applying strains simultaneously and appropriately.

### 5.2.2. Formation of high-density grain boundaries

Grain boundaries (GBs), also described as line defects, play a vital role in tuning the density of states and tailor the HER performance in atomically ultra-thin 2D TMD film. Due to hardship in controlling the structure and density of GBs, the connections of GBs and the HER activity attracts little attention. Theoretically, high-density GBs can create a new band near the Fermi level and narrow the energy gap, thereby modulating the HER spontaneous kinetics, which is similar to the role of vacancies[29].

He et al.[30] fabricated the wafer-size ultra-thin $MoS_2$ film (up to ~$10^{12}$ cm$^{-2}$) via Au-quantum-dot-assisted vapor-phase growth. Its exceedingly superior intrinsic electrocatalytic performance was comprehensively demonstrated by the onset potential (-25 mV) and the Tafel slope (54 mV/dec). The current density of the TMD nanograin film (about 1000 mA/cm$^2$) shows better performance than using CVD film bottom surface, with a single edge and GB assembled. Most importantly, single Au is confirmed to have possibilities to contribute to the overall HER performance as a subsidiary part, but the main contributor is still the $MoS_2$ nanograin film[30].

# 6. Optimising the electronic structure

## 6.1. Heteroatom doping

Doping heteroatoms is an efficient method to control the electronic structure and regulate the HER catalytic behaviour. Both metal sites and non-metal sites can be replaced to alter and tune the basic properties of TMD materials. Common categories of doped atoms can be divided into three parts: noble metals (Pt, Au, and Pd), non-noble metals (Co, Ni, Fe, Cu, V, and Zn) and non-metals (O, N, P, and B)[31]. The doped atoms can be located on edge or in the basal plane.

### 6.1.1. Metal doping

After the addition of metal dopants, due to the difference in the bond lengths and angles of Mo-S and X-S (X=metal) bonds, the in-plane distortion may generate the new electronic states and reduce the energy gap to alter the Gibbs Free Energy. Doping modification greatly optimises the adsoprtion of H atom and thereby, improve the in-plane intrinsic activity[8, 28, 32].

However, different metals may exert different influence on the HER kinetics and activity. Co/Ni-doped TMDs can remarkably reduce the value of $\Delta G_H$ and tune the local density of states[33]. Specifically, according to the DFT calculation, Co/Ni atoms result in a striking decrease regarding the activation on the basel plane with a sharp change of $\Delta G_H$ from 1.5 eV to 0.15 eV under CV and LSV measurement. Conversely, Co/Ni modification causes little effect on the edge S sites.

### 6.1.2. Non-metal doping

Unlike the metal doping, nonmetal-doped TMDs do not only optimise the $\Delta G_H$, but also replace the S sites to generate the lattice disorder and distortion because of the bond length and angle difference. This time, the modified electronic states will appear to modulate the intrinsic band structure and stimulate the electron conductivity as well as the $H_2$ generation reaction.

With an onset overpotential of 120 mV and a Tafel slope of 55 mV/dec, a boosted HER performance got a thorough verification by incorporating oxygen atoms, which reveals the tuning in both covalent and conduction band of TMDs[30].

## 6.2. Surface functionalisation with organic molecules

The coupling system can decrease the gap width and facilitate the charge transfer process by altering the band structure and TMDs-based hydrogen adsorption kinetics. Organic ligands with specific functional groups are the most common category as an electron donor[34, 35]. As the catalytically organic ligand, the only one for HER enhancement is thiobarbituric acid (TBA) among versatile functional groups. As mentioned in Presolski et al.[35], the main reasons are 1T-phase metallic properties, the poorly basic and wettable environment on the surface. Also, the calculation result shows a highly boosted activity with low or high coverage of TBA molecules, where 50% of the TBA contributes to the maximally superior performance.

However, the tedious processing and time-consuming procedures may hinder the main focus of this functionalised coupling system.

## 7. Summary and perspectives

We comprehensively summarised the modification strategies and the state-of-the-art advances of HER electrocatalysts based on 2D TMDs. Following the composition-characterisation-construction guideline, we offered three methodologies for HER enhancement: 1) to increase the active sites; 2) to improve the intrinsic conductivity and activity; 3) to optimise the electronic structure. These strategies can boost HER performance individually or in a synergic way to highlight their roles in structural design and electronic modulation. Both theoretical and experimental findings play vital roles in more insight into TMDs-related HER system, as comprehensively summarised in Fig. 8.

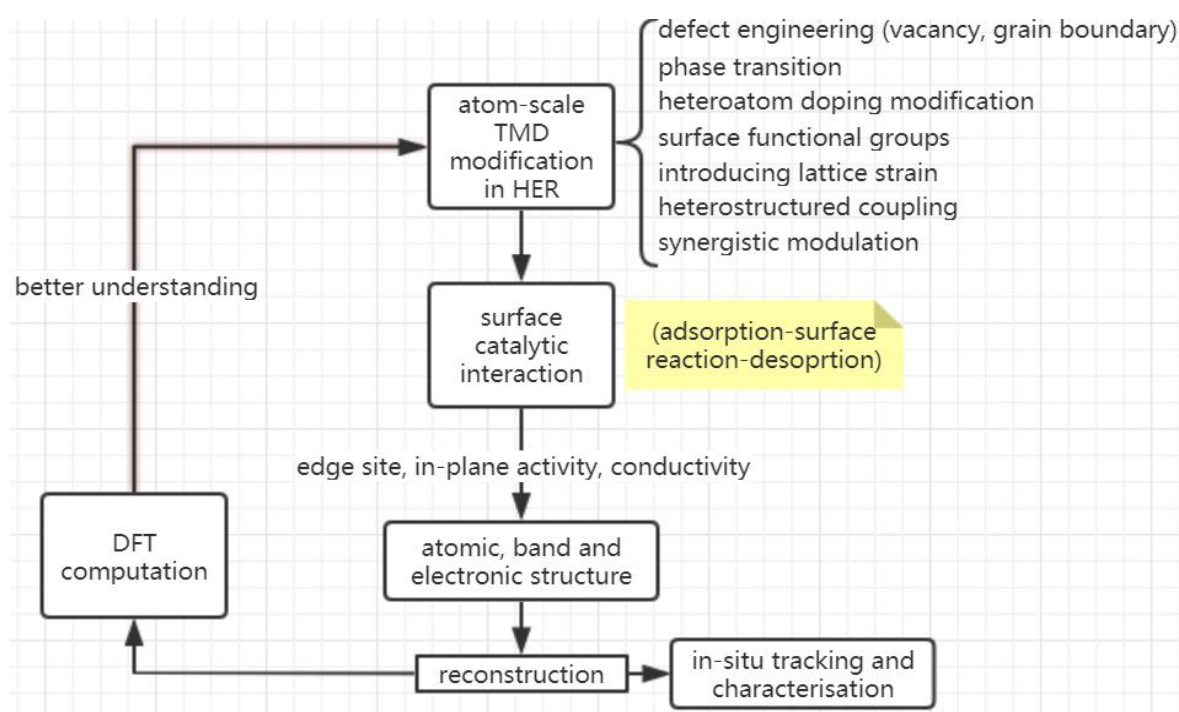

**Fig. 8** schematic principles of the optimal design and modulation of TMDs-based HER electrocatalysts based on the composition-characterisation-construction guideline

However, there is still a long way to go before the broad application of TMDs-based catalysts in water electrocatalysis:
1) Regarding the nano-level synthesis of TMDs, there is a lack of systematic theoretical guidance and well-tuned fabrication methods;
2) The correlations in HER catalytic activity and nanostructures of TMDs is unclear;
3) More intelligent algorithms are urgently needed to narrow the gap between experimental and simulated results.
4) Lastly, the long-term stability of catalysts should be highlighted. The large-scale application needs electrocatalysts with extraordinary long-term stability and durability.

Overall, the 2D TMDs exhibit great potential to replace the noble-metal HER electrocatalysts

(Pt) for efficient water electrochemical splitting. By the rational optimal design of TMDs. It is possible to achieve a wide-ranging commercial application.